# Structural, magnetic and insulator-to-metal transitions under pressure in the GaV$_4$S$_8$ Mott insulator:
# A rich phase diagram up to 14.7 GPa

J. Mokdad[1-2]*, G. Knebel[1-2] C. Marin[1-2], J.-P. Brison[1-2], V. Ta Phuoc[3], R. Sopracase[3], C. Colin[4] and D. Braithwaite[1-2]

*1 Univ. Grenoble Alpes, Grenoble, France*
*2 CEA, IRIG-PHELIQS, F-38000 Grenoble, France*
*3 GREMAN-UMR 7347 CNRS - Université de Tours, Parc de Grandmont, 37200 Tours, France*
*4 Univ. Grenoble Alpes, CNRS, Institut Néel, 38000 Grenoble, France*

In addition to its promising potential for applications, GaV$_4$S$_8$ shows very interesting physical properties with temperature and magnetic field. These properties can be tuned by applying hydrostatic pressure in order to reveal and understand the physics of these materials. Not only pressure induces an insulator-to-metal transition in GaV$_4$S$_8$ but it also has an interesting effect on the structural and magnetic transitions. Using a combination of AC calorimetry, capacitance, and resistivity measurements under pressure, we determine the evolution of the structural and magnetic transitions with pressure and thus establish the T-P phase diagram of GaV$_4$S$_8$. To detect the insulator-to-metal transition, we use optical conductivity and DC resistivity measurements and we follow the evolution of the Mott gap under pressure. The structural transition temperature increases with pressure and a second transition appears above 6 GPa indicating a possible new phase with a very small gap. Pressure has surprisingly a very weak effect on the ferromagnetic transition that persists even very close to the IMT that occurs at around 14 GPa, implying that the metallic state may also be magnetic.

## I. INTRODUCTION

GaV$_4$S$_8$ is a lacunar spinel Mott insulator that has gained a lot of attention in the past few years due to its potential applications in Mott memories and neuromorphic computing in addition to its interesting physical properties and rich phase diagram at low temperatures and magnetic fields [1]. The lacunar spinels are a special kind of Mott insulators due to the fact that the electrons are highly localized within metal clusters (see Fig. 1) and not localized on atomic sites like in a usual Mott insulator. It is the long inter-cluster distances of about 4 Å [3] that keeps the electrons from circulating and thus makes the material insulating. Many studies exist on GaV$_4$S$_8$ at ambient pressure [2][3][4] showing a ferroelectric Jahn-Teller symmetry-lowering structural transition at 43.3 K from cubic $F\bar{4}3m$ to rhombohedral $R3m$ and a transition to a ferromagnetic state at 12.7 K, which contains skyrmionic and cycloidal phases at low temperature and low magnetic field.

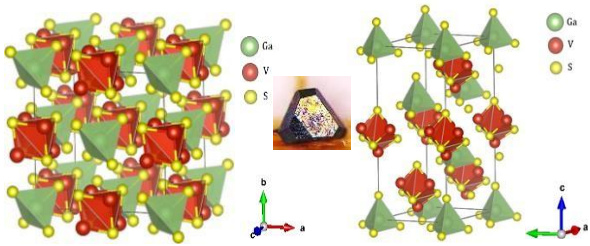

Figure 1: The cubic lattice structure of the lacunar spinel GaV4S8 at 300 K (left) with a $F\bar{4}3m$ symmetry and a lattice parameter of a=9.661 Å [11]. The tetrahedral (GaS$_4$)$^{5-}$ cluster and the cubane (V$_4$S$_4$)$^{5+}$ clusters are presented in green and red respectively. At Ts=43.3 K, a Jahn-Teller transition causes an elongation in the V$_4$ and GaS$_4$ tetrahedra which reduces the symmetry to rhomboedric R3m (right). The inset shows a crystal obtained by the vapor-phase transport growth with a large triangular face perpendicular to the [111] crystallographic direction.

Therefore, the application of hydrostatic pressure is a very pertinent tool in the lacunar spinel family since it decreases the inter-cluster distance and hence increases the transfer interaction while maintaining the same lattice structure [6]. In other words, by forcing the clusters closer together, the material could become conducting by hopping of the electrons between the clusters. By this procedure, pressure can induce Mott insulator-to-metal transitions. Several electrical transport studies exist on these materials showing a transition to a metallic phase and even the emergence of superconductivity in some members of the family [7] although this study was not performed on GaV$_4$S$_8$. Moreover, these resistivity studies were not able to explore inside the insulating state, so the high pressure/low-temperature phase diagrams of these systems are unknown. We initiated this study, to reveal the insulator to metal transition in GaV$_4$S$_8$ and also to establish the effect of pressure on the rich low-temperature phase diagram. To achieve this we combined specific heat, capacitance, and resistivity measurements, under pressure up to 14.7 GPa in a diamond anvil cell with in-situ pressure tuning. We have determined the evolution under pressure of the temperature of the structural and magnetic transitions and thus established the detailed T-P phase diagram. We also follow the evolution of the gap energy with pressure and reveal the insulator-to-metal transition, both with transport and optical conductivity experiments. We find that the structural transition temperature T$_S$ increases with pressure and we detect a second anomaly at high pressure. We find two different insulating or semiconducting regimes at temperatures above and below the structural transition, and above a pressure of about 10 GPa the low-temperature state is a small gap semiconductor. The system becomes fully metallic at around 14 GPa. The magnetic transition temperature stays relatively stable with pressure and magnetism persists at least up to 12 GPa close to the insulator-to-metal transition (IMT) at around 14 GPa. No sign of superconductivity was observed.

## II. EXPERIMENTAL DETAILS

Single crystals of $GaV_4S_8$ were grown using vapor-phase transport method in a multi-zone tube furnace with iodine as a transport agent. The crystals were characterized and oriented by x-ray Laue diffraction. The [111] crystallographic direction was determined by Laue diffraction as perpendicular to the large triangular surface of the crystal (see Fig. 1). Specific heat and magnetization measurements were made in commercial devices (Quantum Design PPMS and MPMS). The ambient pressure phase diagram versus temperature and magnetic field was determined from the magnetization measurements as well as pyrocurrent measurements, performed using a Keithley Electrometer as a function of temperature between 2 K and 100 K and fields between 0 and 200 mT. These measurements were carried out on relatively large single crystals.

Measurements under hydrostatic pressure (except for optical conductivity) were performed in a diamond anvil cell with liquid argon as a pressure transmitting medium. For these experiments, samples were polished to a thickness of about 30-40 μm perpendicular to [111], then cut to fit into the pressure chamber. All the samples used for measurements under pressure have typical dimensions of around 200 x 250 x 40 μm$^3$. Pressure was changed using a newly developed in-situ pressure tuning system that relies on a mechanical screw device. In contrast to a previous system using helium bellows, this system can be used over the whole temperature range 2 K-300 K. The pressure was determined in situ by the ruby fluorescence technique. The low-temperature measurements were carried out in a helium flow cryostat allowing to cool the sample down to 2.3 K and a magnetic field up to 9 T can be applied.

AC calorimetry measurements under pressure were done in the DAC on a 150 x 250 x 30 μm$^3$ $GaV_4S_8$ sample following the procedure described by Demuer et al. [15]. The sample was heated optically with a laser chopped at a frequency of 1 kHz. An AuFe/Au thermocouple was attached to the sample. The amplitude and phase shift of the temperature oscillations were measured with lock-in detection. For dielectric measurements, a thin layer of gold was evaporated onto both sides of the sample to which wires were attached. We used an Andeen-Hagerling AH 2550A 1kHz capacitance bridge that allows us to measure the capacitance C (pF) proportional to the dielectric constant of the sample and the loss (nS) that is inversely proportional to the resistance. The bridge can measure correctly when the resistance of the sample is of the order of 10 kΩ or higher. For measurements at high temperature or high pressure when the sample was less resistive than this, we performed 2 lead DC resistivity measurements using a current source and nano voltmeter. The total resistance of the current leads and contacts was about 60 Ω and varied by less than 10% over the whole temperature range. Thermal hysteresis between warming and cooling was less than 1.5% coming mainly from thermal lag between the sample and the thermometer. No significant intrinsic hysteresis or memory effects were observed.

Room-temperature pressure-dependent optical conductivity spectra have been deduced from nearly normal incidence sample-diamond reflectivity $R_{sd}(\omega)$ between 550 and 12000 cm$^{-1}$ wavenumbers (8000 cm$^{-1}$ = 1 eV) on a 100 x 100 μm$^2$ single crystal from 0 to 17.5 GPa using a BETSA membrane diamond anvil cell. The sample was loaded inside a gasket hole together with KBr as a hydrostatic medium. The gasket was used as a reference mirror. $R_{sd}(\omega)$ was measured by using a homemade high-vacuum microscope including an X15 Schwarzchild objective connected to a BRUKER IFS 66v/S Fourier Transform Spectrometer with a Mercury-Cadmium-Telluride detector, Glowbar, and tungsten light sources. Pressure was measured with the standard ruby fluorescence technique. In order to obtain the optical conductivity from the $R_{sd}(\omega)$, we used a Variational Dielectric Function method[12].

## III. RESULTS

### A. Phase diagram of $GaV_4S_8$ at ambient pressure

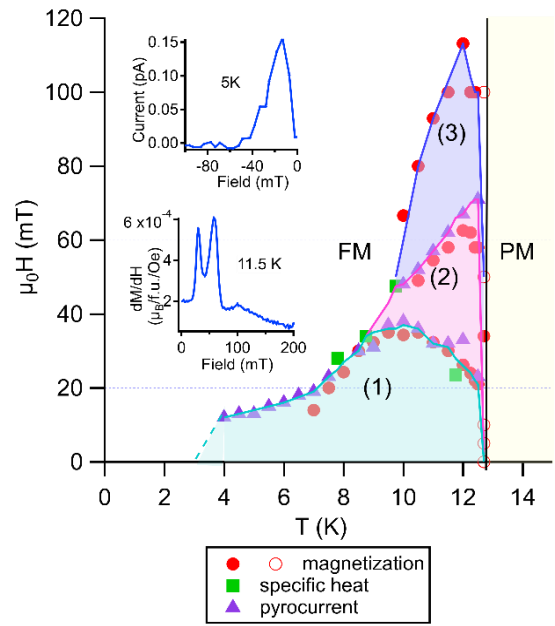

Figure 2: Phase diagram of $GaV_4S_8$ at ambient pressure. The phase boundaries as determined by magnetization, specific heat and pyrocurrent measurements are indicated by circles, squares, and triangles respectively. Three magnetic phases indicated as (1), (2) and (3) correspond to cycloidal, skyrmionic, and skyrmionic* phases respectively by comparison with the phase diagram published by Kezsmarki et al [3]. The first inset shows the anomaly at 13 mT in the pyrocurrent (pA) at 5 K. The curve has been shifted by 13 mT so that the curves of the positive and negative field sweep are symmetric at zero field. The second inset presents the field-derivative of magnetization versus field at 11.5 K: three anomalies appear at 30 mT, 58 mT and 100 mT indicating the transitions between the three magnetic phases.

At $T_m$=12.7 K, GaV$_4$S$_8$ undergoes a magnetic transition that was first thought to be an ordinary ferromagnetic transition [3] [8]. Later it was found that in fact, at 12.7 K, the material enters a cycloidal phase before becoming a normal ferromagnet at lower temperatures. The phase diagram becomes even richer when applying magnetic field in the <111> direction, with different magnetic phases emerging up to 110 mT [1] [2] [3].

Figure 2 presents the phase diagram obtained on our single crystals of GaV$_4$S$_8$ at ambient pressure based on magnetization, pyrocurrent and specific heat measurements with a magnetic field applied along the <111> direction. The nature of the different phases has been identified by comparison to other published phase diagrams [1] [2] [3]. The field-derivative of the magnetization, with B // <111>, and the pyrocurrent both show sharp peaks corresponding to the transitions between the different magnetic phases (see inset of Fig.2). Our results are in quite good agreement with the previous studies. With pyrocurrent and magnetization measurements we detect two magnetic states corresponding to the reported cycloidal and skyrmionic phases that persist up to 20 mT and 70 mT respectively at 12.5 K with the magnetic field parallel to the [111] direction. However, the pyrocurrent measurement shows a clear anomaly at 13 mT at 5 K (see inset Fig.2) that persists down to 4 K and only disappears completely at 2 K, implying that the cycloidal phase at zero field persists to temperatures down to 3 K before becoming ferromagnetic.

### B. Measurements under pressure

#### 1) AC Calorimetry measurements:

In the AC calorimetry measurements (Figure 3), we found that the phase shift ΔΘ of the signal is more sensitive to the phase transitions than the temperature oscillation amplitude $T_{ac}$. The magnetic transition appears as a well-defined peak in the phase of the signal, and the aspect of the transition remains almost unchanged over the whole pressure range explored here, up to 12.5 GPa. The magnetic transition temperature $T_m$ decreases slightly with pressure from 12.7 K at ambient pressure, passes by a minimum of $T_m$=9.6 K at 6 GPa then increases again to reach $T_m$=15 K at 12.5 GPa. So surprisingly pressure has a rather weak effect on the magnetic transition in GaV$_4$S$_8$, over a pressure range of 12.5 GPa. (See phase diagram in Fig.9)

The structural transition is more difficult to track under pressure with ac calorimetry. Although in the ambient pressure specific heat measurement it appears as a large anomaly, being a first-order transition, the anomaly can be hugely attenuated in the AC measurement. At low pressure it starts out as a small peak that gets smeared out when pressure increases and becomes nearly invisible above 6.2 GPa.

As an overall behavior, the temperature of the structural transition Ts begins at 43.3 K at ambient pressure and increases up to 55 K at 7 GPa.

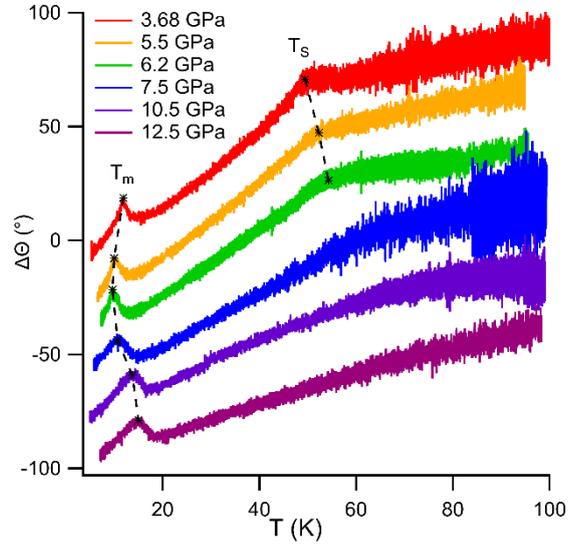

Figure 3: Temperature dependence of the phase ΔΘ in AC calorimetry measurements in GaV$_4$S$_8$ for different pressures. The curves are shifted vertically for clarity. The magnetic and structural transitions are marked by asterisks and the dashed lines are guide to the eye to show the variation of $T_m$ and $T_S$ with pressure. For pressures above 7.5 GPa, the structural transition becomes nearly invisible.

Since the magnetic transition stays relatively stable under pressure and the structural transition anomaly is rather weak, AC calorimetry does not give any indication of the expected insulator-to-metal transition. Therefore we performed dielectric measurements using a capacitance bridge which allows measuring the dielectric constant and the resistivity simultaneously.

#### 2) Dielectric measurements:

Figure 4 presents the temperature dependence of the capacitance (pF) and loss (nS) measured simultaneously by the capacitance bridge for different pressures. At 1.45 GPa the capacitance bridge only measures below 120K when the resistance of the sample is greater than 6.7 kΩ. In contrast to the AC calorimetry measurement, at pressures up to 6 GPa, the structural transition shows up here as a huge anomaly in both the capacitance and the loss. We take the maximum of the peak in capacitance and loss curves as a criterion to determine the temperature of the structural transition (see Fig.4).

The transition is clearly visible in both the capacitance and the loss up to 6 GPa and the transition temperature $T_S$ initially increases with pressure. At higher pressure the anomaly in the loss disappears. In the capacitance measurements above 6.9 GPa, the peak that seems to correspond to $T_S$ is transformed into a wide bump that shifts to lower temperatures and another marked peak that we will mark as $T_X$ appears at higher temperature. The capacitance bridge stops measuring at pressures above 9.2 GPa when the loss becomes too high to measure (electrical resistance too low) and the capacitance becomes too low.

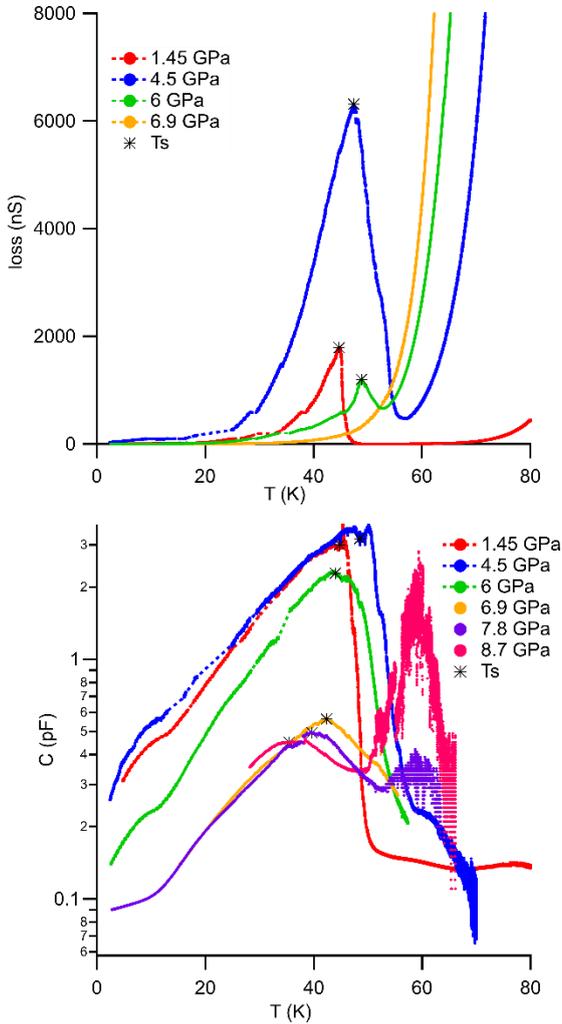

Figure 4: (Top graph) Loss versus temperature under pressure. The loss shows a peak for the structural transition $T_S$. It is marked by asterisks on the curves. $T_S$ increases with pressure and becomes invisible at 6.9 GPa. (Bottom graph) Capacitance versus temperature under pressure showing one peak for the structural transition $T_S$ for pressures below 7.8 GPa. Above this pressure, Ts gets smeared out and another peaks Tx appears at higher temperatures. Ts has been marked by asterisks on the curves.

### 3) DC Resistivity measurements:

Two-lead DC resistivity measurements were performed in the pressure/temperature ranges where the sample resistance was too low to measure with the capacitance bridge, and to some extent for intermediate resistance values where both types of measurement were carried out. This technique allowed values of resistance up to about 1 MΩ to be measured reliably. Figure 5 shows the temperature dependence of the electrical resistance of the sample in the temperature range below 100 K as a function of pressure for $GaV_4S_8$. At low pressures, the resistance shows a semi-conducting like behavior with the structural transition clearly visible at 3.2 GPa. The transition gets smeared out with increasing pressure which makes it difficult to determine the exact value of Ts up to 9.2 GPa. However above this pressure, as the sample becomes less semi-conducting the transition actually becomes more pronounced again.

At 13.7 GPa, the resistance of the sample decreases with decreasing temperature except at very low temperatures where a semi-conducting behavior is still found. A metallic-like behavior appears at 14.7 GPa over the whole temperature range.

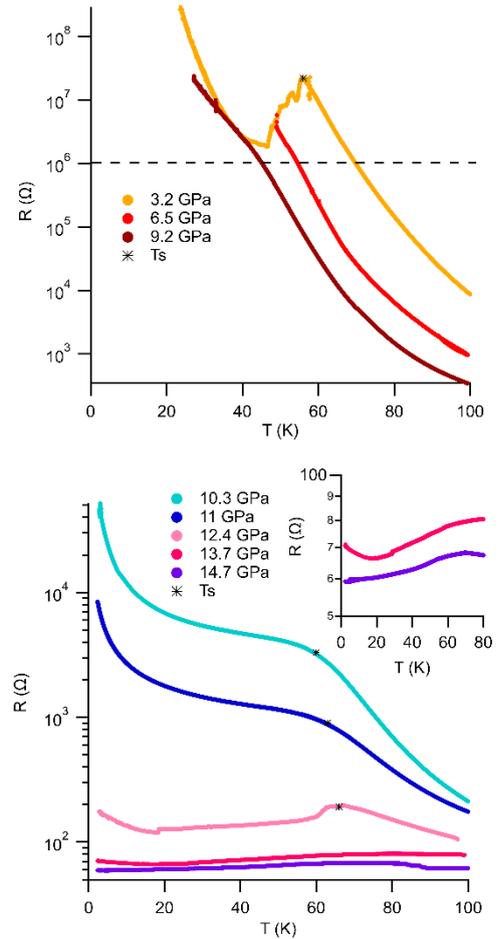

Figure 5: Temperature dependence of the DC resistance in $GaV_4S_8$ for different pressures. The structural transition is marked by an asterisk. Ts has been determined from the intersection of the two lines drawn tangent to the resistance curves R(T) around the anomaly. The inset shows the resistance vs temperature for 13.7 GPa and 14.7 GPa showing an insulator-to-metal transition. A dashed line at 1 MΩ indicates the value above which the measured resistance becomes unreliable.

### 4) Reflectivity measurements:

Pressure-dependent reflectivity spectra at the sample-diamond interface $R_{sd}(\omega)$ are shown in Fig. 6. Strong two-photon diamond absorption prevents the measurement of $R_{sd}$ between 1750 cm$^{-1}$ and 2700 cm$^{-1}$.

On increasing the pressure from 0 to 17.5 GPa, the reflectivity is progressively enhanced at low frequency. Up to 5 GPa, Fabry-Perot interferences can be observed below 2000 cm$^{-1}$ suggesting that the sample behaves like a transparent layer in this frequency range, confirming its insulating nature.

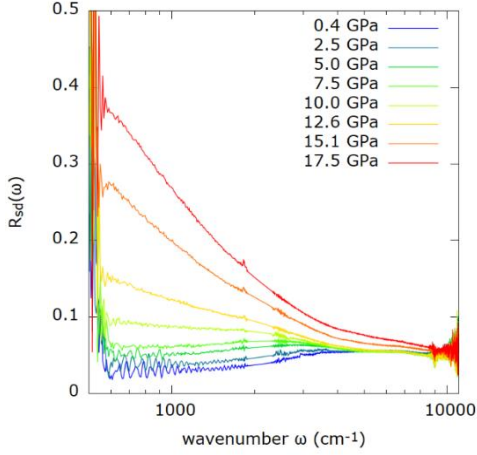

Figure 6: Pressure dependence of the reflectivity spectra at the sample-diamond interface.

As pressure increases, the fringes disappear, meaning a decrease of the skin depth and the existence of an electronic background at low frequency.

Those results are consistent with an insulator-to-metal transition under pressure. Quantitative information will be extracted from the optical conductivity spectra deduced from $R_{sd}$ (see next section).

### C. Insulator-to-metal transition IMT in GaV$_4$S$_8$ under pressure

a. Transport measurements

Another way to determine the insulator-to-metal transition IMT is to follow the variation of the gap energy with pressure. The exponential increase of resistivity to lower temperatures indicates thermal activation of carriers. Therefore, we can extract the gap energy from an Arrhenius law:

$$\sigma = \sigma_0\, e^{-\frac{E_G}{2k_B T}}$$

Figure 7a shows the Arrhenius plots for different pressures. At high temperatures above $T_S$, the thermal activation law is valid and $\ln \rho\ (T^{-1})$ is linear over a wide temperature range up to 8 GPa. Above this pressure the behavior becomes less linear therefore the estimation of the gap energy has been done on a smaller temperature range. Below the structural transition (25K<T<45K), an activation law is still valid with a much smaller gap.

Figure 7b shows the pressure dependence of the gap above and below the structural transition. The gap energy determined above the structural transition from resistivity measurements decreases linearly with pressure and by extrapolation a pressure of 14 GPa is found for the IMT in good agreement with the global behavior over the whole temperature range found in the DC resistivity. Studying the pressure dependence of the optical gap at 300 K (see below), one can observe its linearity, similarly to the gap obtained from the Arrhenius law around 100 K. The pressure of the IMT obtained by the former technique is 2-3 GPa smaller than the latter.

Below the structural transition we find a semiconducting behavior with a gap of 60 meV, in agreement with low temperature optical conductivity [13]. Up to 6 GPa this gap decreases with pressure almost linearly. Above this pressure, $E_G$ increases to reach a maximum at 8 GPa then decreases again to a very small value, reaching 2.3 meV at 11 GPa.

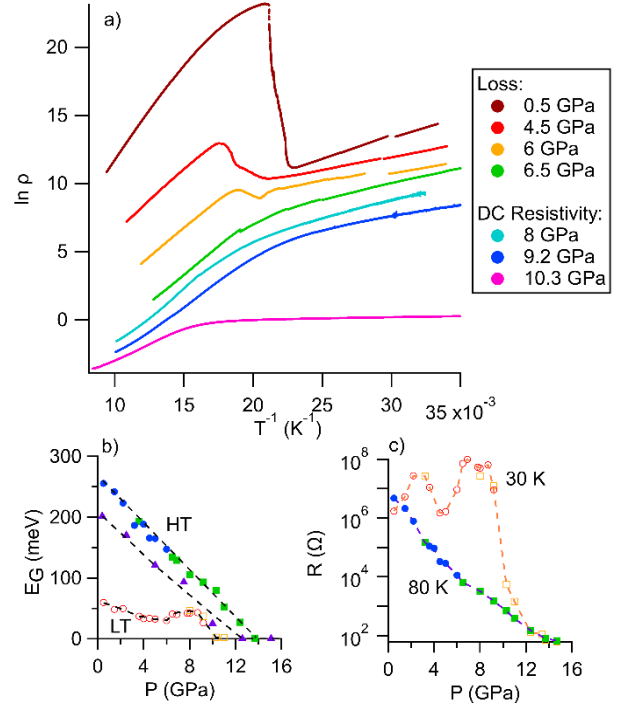

Figure 7: a) $\ln \rho$ versus $T^{-1}$ showing a linear behavior at high and low temperatures up to 6.5 GPa. The curves are shifted vertically for clarity. b) Pressure dependence of the gap energy $E_G$ above and below the structural transition (close and open symbols respectively) computed from loss, DC resistivity measurements and optical conductivity indicated by circles, squares and triangles respectively. The pressure dependence of the optical gap is obtained at room temperature by a standard extrapolation technique. c) Pressure dependence of the resistance at 80K and 30K (close and open symbols respectively) computed from loss and DC resistivity measurements indicated by circles and squares respectively.

Figure 7c shows a smooth decrease of the resistance at 80K up to 14.7 GPa. On the other hand, at 30K, the resistance is relatively stable up to 9.2 GPa then suddenly decreases by several orders of magnitude (see also Figure 5) to reach 60Ω at 14.7 GPa which represents mostly the resistance of the coaxial wires, indicating also the IMT.

b. Optical conductivity

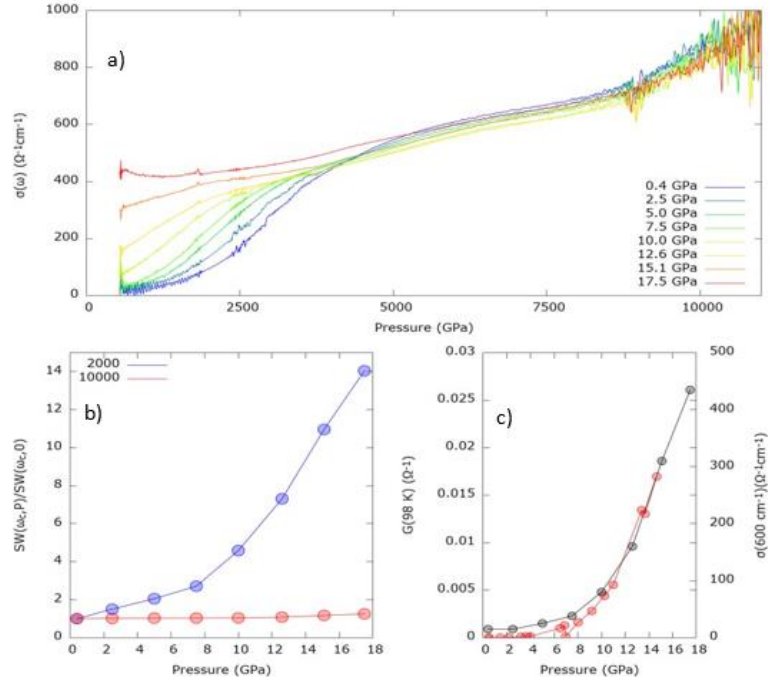

Figure 8: a) Pressure dependence of the optical conductivity at room temperature. b) Pressure dependence of the normalized Spectral Weight (SW) on the whole spectrum (red) and at low energy (blue). c) Pressure dependence of the lowest measured optical conductivity (gray), pressure dependence of the conductance (red) (obtained from the transport measurements).

The pressure-dependent optical conductivities $\sigma(\omega)$ deduced from the $R_{sd}$ at 300 K are shown in Fig. 8.a. The ambient pressure $\sigma(\omega)$ consists of a broad mid-infrared band centered around 0.6 eV (5000 cm$^{-1}$) and the tail of higher energy excitations consistently with Reschke et al. [13]. An optical gap of 200 meV (1600 cm$^{-1}$) can be extracted by the standard extrapolation procedure, consistently with the 250 meV obtained from transport measurements (see Figure 7b). Note that the gap is substantially larger than for isostructural GaTa$_4$Se$_8$ [14] (120 meV).

With increasing pressure, the optical gap progressively closes up to 10 GPa. Between 10 and 12.6 GPa, the system undergoes an IMT and the extrapolation of $\sigma(\omega)$ at low-frequency $\sigma_{opt}$ increases with pressure. At the highest pressure, the onset of a narrow Drude peak, characteristic of a metallic behavior is observed.

Figure 8.c depicts the pressure evolution of the lowest measured $\sigma(\omega)$ ($\sigma(600$ cm$^{-1})$) at 300 K in the Insulating Paramagnetic Paraelectric phase and the conductance measured by transport at 98 K in the same thermodynamic phase. Both curves are very similar, they both show a steep increase at some pressure which is the signature of the IMT.

The spectral weight (SW) defined as:

$$SW_{\Omega_0}^{\Omega}(P) = \int_{\Omega_0}^{\Omega} \sigma(\omega, P) d\omega$$

is proportional to the charge carrier density involved in the [$\Omega_0$, $\Omega$] frequency range.

In the insulating phase, when P increases, some spectral weight is transferred from high to low energy, filling the gap region and evidencing an isosbestic point around 0.5 eV (4000 cm$^{-1}$) in the optical conductivity spectra. From 12.6 to 17.5 GPa, there is no more spectral weight transfer as the whole spectrum increases with pressure. To be more quantitative, Figure 8.b shows the total measured SW (up to 10000 cm$^{-1}$) and the low frequency SW (up to 2000 cm$^{-1}$). Total SW increases by ~20% from lowest to highest pressure meaning an almost SW conservation below 1.25 eV. On the contrary, low frequency SW drastically increases with pressure. Such a SW transfer over a large energy scale under pressure is a usual fingerprint of strongly correlated systems [14].

### D. Phase diagram of $GaV_4S_8$ under pressure

Figure 9 shows the phase diagram of $GaV_4S_8$ under pressure up to 14.7 GPa determined from ac calorimetry, capacitance and resistivity measurements under pressure.

Considering the overall behavior of the transitions, we can see that $T_S$ increases when applying pressure from 43.3K at ambient pressure to about 55K at 6 GPa. At higher pressures two anomalies are seen, one (labeled $T_X$ as a second peak in the capacitance and a kink in the DC resistance measurements) continues to shift to higher temperatures, and a second one, that seems to be the continuation of $T_S$ in the capacitance measurements that moves to lower temperatures. The shape of the anomaly in the ac calorimetry is stable over the whole pressure range, and the magnetic transition temperature slightly changes little with pressure, although a minimum is seen around 6 GPa.

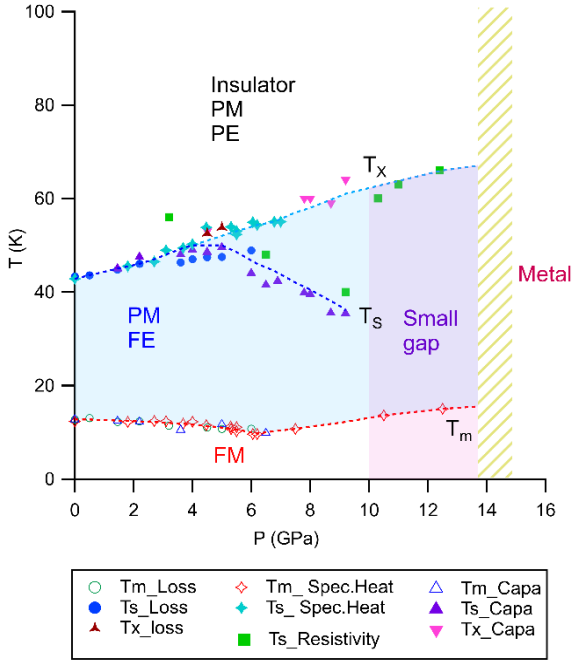

Figure 9: T-P phase diagram of $GaV_4S_8$ determined by specific heat, capacitance and resistivity measurements under pressure up to 14.7 GPa. Dotted lines are guide to the eye.

### IV. DISCUSSION

In this T-P phase diagram, one could clearly note three different insulating/semiconducting regimes.

1) At ambient pressure, the material is a Mott insulator in the cubic high-temperature phase with a Mott gap of 260 meV and becomes semiconducting below the structural transition with a 60 meV gap. The gap below the structural transition becomes smaller when the pressure increases up to 6 GPa. It is interesting to mention that some theoretical studies [5] [10] predict that the structural distortion might open a gap at the Fermi level even without strong electronic correlations, whereas we find actually a smaller gap in the low temperature phase.

2) Between 6 GPa and 10 GPa, an additional phase transition appears. Further studies are necessary to determine its nature. We have noticed that in this pressure range the initial structural transition Ts becomes less marked. This is seen in AC calorimetry measurements where Ts becomes invisible, the peak in resistivity disappears and the peak in the capacitance gets smeared out. The gap energy below the structural transition becomes larger above 6 GPa and the conduction regime starts to deviate from the thermal activation law.

3) The Mott gap determined from the high-temperature resistivity decreases linearly with pressure, reaching zero at around 14.2 GPa ± 0.4 GPa, and at this pressure $GaV_4S_8$ Becomes fully metallic. However above 10 GPa the low temperature gap becomes very small and the absolute value of the resistance decreases by several orders of magnitude at low temperatures (see Figure 7). So in quite a large P-T range the system is a rather small gap semiconductor. These results are confirmed by optical conductivity at 300K where a closing of the gap at around 12.5 GPa is found. Note that other members of the lacunar spinel family undergo this IMT at close pressures: 15 GPa for $GaTa_4Se_8$ and 19 GPa for $GaNb_4Se_8$ [7]. In $GaV_4S_8$ the anomaly due to the magnetic transition is clearly visible in AC calorimetry measurements up to 12.5 GPa, although we can't rule out a change in the type of magnetic order. So the small gap phase is almost certainly magnetic and possibly the high pressure metallic phase may be too, so at low temperature $GaV_4S_8$ may switch from a ferromagnetic insulator to a ferromagnetic metal. This has been seen in other Mott insulators like $Ca_2RuO_4$[9] where the IMT happens with the material staying magnetic.

Even in the metallic state we found no superconductivity in $GaV_4S_8$, whereas it has been seen in other members of the spinel family like $GaTa_4Se_8$ and $GaNb_4Se_8$ even slightly before the IMT [7]. Nevertheless these compounds do not exhibit structural transitions and remain in the cubic phase over a broad temperature range [7]. They also show very weak magnetism and do not exhibit magnetic ordering down to 1.6K [5], [7], contrarily to $GaV_4S_8$ that is ferromagnetic below $T_m$.

### V. CONCLUSION

$GaV_4S_8$ presents a rich T-P phase diagram with a structural transition increasing to higher temperatures with pressure and the emergence of a second transition at high pressure. To our knowledge, this would be the first phase diagram showing the evolution of the structural and magnetic transitions under pressure in a lacunar spinel. Supplementary measurements are required to follow the nature of the structural changes in $GaV_4S_8$ under pressure and determining the magnetic properties of the different phases and the nature of the new phase limited by $T_X$. We observed a pressure-induced insulator-to-metal transition between 10 and 12.6 GPa at 300K that shifts to 14 GPa at low temperatures. Interestingly, the Mott gap in the cubic high temperature phase decreases linearly with pressure and closes at the IMT. On the other hand, a smaller gap at low temperatures closes at a lower pressure. The material stays ferromagnetic at least up to the IMT and further measurements are needed to determine if it stays ferromagnetic in the metallic phase as well.

## VI. ACKNOWLEDGEMENTS:

DB acknowledges the contributions of Andrew Huxley and Paul Attfield to implement capacitive measurements in the diamond anvil cell. J.M. acknowledges financial support from the CEA "Amont-Aval" PhD program.

## VII. REFERENCES:

[1] I. Kézsmárki, S. Bordács, P. Milde, E. Neuber, L. M. Eng, J. S. White, H. M. Rønnow, C. D. Dewhurst, M. Mochizuki, K. Yanai, H. Nakamura, D. Ehlers, V. Tsurkan, and A. Loidl, Néel-type skyrmion lattice with confined orientation in the polar magnetic semiconductor GaV4S8, Nat. Mater. **14**, 1116 (2015).

[2] E. Ruff, S. Widmann, P. Lunkenheimer, V. Tsurkan, S. Bordács, I. Kézsmárki, and A. Loidl, Multiferroicity and skyrmions carrying electric polarization in GaV4S8, Sci. Adv. **1**, e1500916 (2015).

[3] R. Pocha, D. Johrendt, and R. Pöttgen, Electronic and structural instabilities in GaV4S8 and GaMo4S8, Chem. Mater. **12**, 2882 (2000).

[4] S. Widmann, E. Ruff, A. Günther, H.-A. K. von Nidda, P. Lunkenheimer, V. Tsurkan, S. Bordács, I. Kézsmárki, and A. Loidl, On the multiferroic skyrmion-host GaV4S8, Philos. Mag. **97**, 3428 (2017).

[5] Y. Wang, D. Puggioni, and J.M. Rondinelli, Assessing exchange-correlation functional performance in the chalcogenide lacunar spinels $GaM_4Q_8$ (M= Mo, V, Nb, Ta; Q= S, Se), arXiv preprint arXiv: 1905.09170 (2019).

[6] M. Imada, A. Fujimori, and Y. Tokura, Metal-insulator transitions, Reviews of modern physics **70**, 1039 (1998).

[7] M.M. Abd-Elmeguid, B. Ni, D.I. Khomskii, R. Pocha, D. Johrendt, X. Wang, and K. Syassen, Transition from Mott Insulator to Superconductor in GaNb4Se8 and GaTa4Se8 under High Pressure, Phys. Rev. Lett. **93**, 126403 (2004).

[8] C.S. Yadav, A.K. Nigam, and A.K. Rastogi, Thermodynamic properties of ferromagnetic Mott-insulator GaV4S8, Physica B: Condens. Matter **403**, 1474 (2008).

[9] P.L. Alireza, F. Nakamura, S.K. Goh, Y. Maeno, S. Nakatsuji, Y.T. Chris Ko, M. Sutherland, S. Julian and G.G. Lonzarich, Evidence of superconductivity on the border of quasi-2D ferromagnetism in Ca2RuO4 at high pressure, Journal of Physics: Condens. Matter **22**, 052202 (2010).

[10] M. Sieberer, S. Turnovszky, J. Redinger, and P. Mohn, Importance of cluster distortions in the tetrahedral cluster compounds GaM4X8 (M= Mo, V, Nb, Ta; X= S, Se): Ab initio investigations, Phys. Rev. B **76**, 214106 (2007).

[11] Y. Sahoo, and A.K.Rastogi, Evidence of hopping conduction in the V4-cluster compound GaV4S8, Journal of Physics: Condens. Matter **5**, 5953 (1993).

[12] A. B. Kuzmenko, Kramers-Kronig constrained variational analysis of optical spectra, Rev. Sci. Instrum. **76**, 083108 (2005).

[13] S. Reschke, F. Mayr, Z. Wang, P. Lunkenheimer, Weiwu Li, D. Szaller, S. Bordács, I. Kézsmárki, V. Tsurkan, and A. Loidl, Optical conductivity in multiferroic GaV4S8 and GeV4S8: Phonons and electronic transitions, Phys. Rev. B **96**, 144302 (2017).

[14] V.T. Phuoc, C. Vaju, B. Corraze, R. Sopracase, A. Perucchi, C. Marini, P. Postorino, M. Chligui, S. Lupi, E. Janod, and L. Cario, Optical conductivity measurements of GaTa4Se8 under high pressure: Evidence of a bandwidth-controlled insulator-to-metal Mott transition, Phys. Rev. Lett. **110**, 037401 (2013).

[15] A. Demuer, C. Marcenat, J. Thomasson, R. Calemczuk, B. Salce, P. Lejay D. Braithwaite, J. Flouquet, Calorimetric study of CeRu2Ge2 under continuously swept hydrostatic pressure up to 8 GPa, Journal of low temperature physics **120**, 245 (2000).